\documentclass{article}
\usepackage{graphicx}
\usepackage{cite}
\begin{document}

\title{\bf Minimal Length and the Quantum Bouncer: A Nonperturbative Study}

\author{Pouria Pedram\thanks{p.pedram@srbiau.ac.ir}\\
       {\small Department of Physics, Science and Research Branch, Islamic
        Azad University, Tehran, Iran}}

\date{\today}

\maketitle \baselineskip 24pt

\begin{abstract}
We present the energy eigenvalues of a quantum bouncer in the
framework of the Generalized (Gravitational) Uncertainty Principle
(GUP) via quantum mechanical and semiclassical schemes. In this
paper, we use two equivalent nonperturbative representations of a
deformed commutation relation in the form $[X,P]=i\hbar(1+\beta
P^2)$ where $\beta$ is the GUP parameter. The new representation is
formally self-adjoint and preserves the ordinary nature of the
position operator. We show that both representations result in the
same modified semiclassical energy spectrum and agrees well with the
quantum mechanical description.
\end{abstract}


\textit{Pacs}: {04.60.-m}

\section{Introduction}
Various proposals for quantum gravity such as string theory, loop
quantum gravity, quantum geometry, and black-hole physics all agree
on the the existence of a minimum measurable length of the order of
the Planck length $\ell_{Pl}=\sqrt{G\hbar/c^3}\approx 10^{-35}m$
where $G$ is Newton's gravitational constant. It can solve the
renormalizability problem of relativity which is related to the
field theoretical ultraviolet divergences in the theory. Indeed,
beyond the Planck energy scale, the effects of gravity are so
important which would result in discreteness of the very spacetime.
However, by introducing a minimal length as an effective cutoff in
the ultraviolet domain, the theory can be renormalizable.

The introduction of this idea in the form of the Generalized
Uncertainty Principle (GUP) has attracted much attention in recent
years and many papers have been appeared in the literature to
address the effects of GUP on various quantum mechanical systems
\cite{12,13,14,15,16,r2,17,18,19,101,102,103,104,105,106,7,Nouicer}.
In doubly special relativity theories, we consider the Planck energy
as an additional invariant beside the velocity of light
\cite{21,22,23,24,3}. A GUP which is consistent with these theories
is also discussed in Refs.~\cite{main,main2,pedram,pedram3,pedram1}.

In this paper, we consider the problem of the quantum bouncer in the
context of the generalized uncertainty principle in the form
$[X,P]=i\hbar(1+\beta P^{2})$ where $\beta$ is the GUP parameter. In
ordinary quantum mechanics, the energy levels of a test particle
which is bouncing elastically and vertically on the Earth's surface
should be quantized as a manifestation of the particle-wave duality.
This effect has been confirmed a few years ago using a high
precision neutron gravitational spectrometer. Using the perturbative
representations, the effects of the minimal length and/or maximal
momentum on the quantum bouncer spectrum are also discussed in
Refs.~\cite{Brau,pedram1,NP}. Here, we use two equivalent
nonperturbative representations which exactly satisfy the modified
commutation relation and find the GUP corrected energy spectrum of
the quantum bouncer. Therefore, we can relax the assumption of the
smallness of the GUP parameter and obtain the solutions for
arbitrary values of $\beta$. We show that the semiclassical results
agree well with the quantum mechanical results even for the low
lying states.

\section{The Generalized Uncertainty Principle}
Let us consider a generalized uncertainty principle which results in
a minimum observable length
\begin{eqnarray}\label{gup}
\Delta X \Delta P \geq \frac{\hbar}{2} \left( 1 +\beta (\Delta P)^2
+\zeta \right),
\end{eqnarray}
where $\beta$ is the GUP parameter and $\zeta$ is a positive
constant that depends on the expectation values of the momentum
operator. We also have $\beta=\beta_0/(M_{Pl} c)^2$ where $M_{Pl}$
is the Planck mass and $\beta_0$ is of order one. It is
straightforward to check that the inequality relation (\ref{gup})
implies the existence of a minimum observable length as $(\Delta
X)_{min}=\hbar\sqrt{\beta}$. In the context of string theory, we can
interpret this length as the length of the string where it is
proportional to the square root of the GUP parameter. In one
dimension, the above uncertainty relation can be obtained from the
following deformed commutation relation
\begin{eqnarray}\label{gupc}
[X,P]=i\hbar(1+\beta P^2),
\end{eqnarray}
where for $\beta=0$ we recover the well-known commutation relation
in ordinary quantum mechanics. Now using Eqs.~(\ref{gup}) and
(\ref{gupc}), we find the explicit form of $\zeta$ as the
expectation value of the momentum operator i.e. $\zeta=\beta\langle
P\rangle^2$. As Kempf, Mangano, and Mann (KMM) have suggested in
their seminal paper, in momentum space representation, we can write
$X$ and $P$ as \cite{Kempf}
\begin{eqnarray}\label{k1}
P &=& p,\\ X &=& i\hbar \left( 1 + \beta
p^2\right)\frac{\partial}{\partial p},\label{k2}
\end{eqnarray}
where $X$ and $P$ are symmetric operators on the dense domain
$S_{\infty}$ with respect to the following scalar product:
\begin{eqnarray}\label{scalar}
\langle\psi|\phi\rangle=\int_{-\infty}^{+\infty}\frac{\mathrm{d}p}{1+\beta
p^2}\psi^{*}(p)\phi(p).
\end{eqnarray}
Also we have
$\displaystyle\int_{-\infty}^{+\infty}\frac{\mathrm{d}p}{1+\beta
p^2}|p\rangle\langle p|=1$ and $\langle p|p'\rangle=\left( 1 + \beta
p^2\right)\delta(p-p')$. With this definition, the commutation
relation (\ref{gupc}) is exactly satisfied. Note that the KMM
representation is not unique and we can use another representation
using appropriate canonical transformations. For instance, consider
the following exact representation
\begin{eqnarray}\label{x0p01}
X &=& x,\\ P &=&
\frac{\tan\left(\sqrt{\beta}p\right)}{\sqrt{\beta}},\label{x0p02}
\end{eqnarray}
where $x$ and $p$ obey the canonical commutation relation
$[x,p]=i\hbar$. Note that, in contrary to the KMM representation,
this representation is formally self-adjoint and preserves the
ordinary nature of the position operator. Moreover, this definition
exactly satisfies the condition $[X,P]=i\hbar(1+\beta P^2)$ and
agrees with the well-known perturbative proposal
\cite{NP,main1,pedram2}
\begin{eqnarray}\label{xp-old1}
X &=& x,\\ P &=& p\left( 1 + \frac{1}{3}\beta\, p^2
\right),\label{xp-old2}
\end{eqnarray}
to the first order of the GUP parameter. Indeed, the two representations
are equivalent and they are related by the following canonical
transformation:
\begin{eqnarray}
X&\rightarrow&\left[1+\arctan^2\left(\sqrt{\beta}P\right)\right]X,\\
P&\rightarrow&\arctan\left(\sqrt{\beta}P\right)/\sqrt{\beta},
\end{eqnarray}
which transforms Eqs.~(\ref{x0p01}) and (\ref{x0p02}) to
Eqs.~(\ref{k1}) and (\ref{k2}) subjected to Eq.~(\ref{gupc}). We can
interpret $P$ and $p$ as follows: $p$ is the momentum operator at
low energies ($p=-i\hbar\partial/\partial{x}$) and $P$ is the
momentum operator at high energies. Obviously, this procedure
affects all Hamiltonians in quantum mechanics. Consider the
following Hamiltonian:
\begin{eqnarray}
H=\frac{P^2}{2m} + V(X),
\end{eqnarray}
which using Eqs.~(\ref{x0p01}) and (\ref{x0p02}) can be written
exactly and perturbatively as
\begin{eqnarray}\label{H0}
H&=& \frac{\tan^2\left(\sqrt{\beta}p\right)}{2\beta m}+V(x), \\
&=&H_0+ \sum_{n=3}^\infty \frac{(-1)^{n-1} 2^{2n} (2^{2n}-1)(2n-1)
B_{2n}}{2m(2n)!} \beta^{n-2} p^{2(n-1)},\label{H}
\end{eqnarray}
where $H_0=p^2/2m + V(x)$ and $B_n$ is the $n$th Bernoulli number.
The corrected terms in the modified Hamiltonian are only momentum
dependent and are proportional to $p^{2(n-1)}$ for $n\geq 3$. In
fact, the presence of these terms leads to a positive shift in the
energy spectrum. In the quantum domain, this Hamiltonian results in
the following generalized Schr\"odinger equation in quasiposition
representation
\begin{eqnarray}
-\frac{\hbar^2}{2m}\frac{\partial^2\psi(x)}{\partial x^2}+
\sum_{n=3}^\infty \alpha_n \hbar^{2(n-1)}\beta^{n-2}
\frac{\partial^{2(n-1)}\psi(x)}{\partial
x^{2(n-1)}}+V(x)\psi(x)=E\,\psi(x),
\end{eqnarray}
where $\alpha_n=2^{2n} (2^{2n}-1)(2n-1) B_{2n}/2m(2n)!$  and the
second term is due to the GUP corrected terms in Eq.~(\ref{H}).

\section{The Quantum Bouncer}
Although the quantum stationary states of matter in the
gravitational field is theoretically possible, the observation of
such quantum effect is very difficult experimentally. This is due to
the fact that the gravitational quantum effects are negligible for
macroscopic objects and for charged particles the electromagnetic
interaction has the dominate role. So we should use neutral
elementary particles with a long lifetime such as neutrons. A few
years ago, a famous measurement is performed by Nesvizhevsky et al.
using a high precision neutron gravitational spectrometer
\cite{291,292,293}. In this experiment, the quantization of energy
of ultra cold neutrons bouncing above a mirror in the Earth's
gravitational field was demonstrated.

\subsection{Exact solutions}
Here, we are interested to study the effects of the minimal length
on the energy spectrum of a particle of mass $m$ which is bouncing
vertically and elastically on a reflecting hard floor so that
\begin{eqnarray}
V(X)=\left\{
\begin{array}{cc}
mgX&\mathrm{for}\quad X>0,\\\\ \infty\, &\mathrm{for}\quad X\leq 0,
\end{array}
\right.
\end{eqnarray}
where $g$ is the acceleration in the Earth's gravitational field. In
ordinary quantum mechanics ($\beta=0$), this problem is exactly
solvable and the eigenfunctions can be written in the form of the
Airy functions. Moreover, the energy eigenvalues are related to the
zeros of the Airy function. In the semiclassical approximation, the
energy eigenvalues can be obtained using the Bohr-Sommerfeld formula
as
\begin{eqnarray}\label{BS}
E_n\simeq\left(\frac{9m}{8}\left[\pi \hbar
g\left(n-\frac{1}{4}\right)\right]^2\right)^{1/3},
\end{eqnarray}
which is valid with a high degree of accuracy of $\sim1\%$ even for
the lowest quantum state. However, for $\beta\ne0$ the situation is
quite different.

\begin{figure}
\begin{center}
\includegraphics[width=8cm]{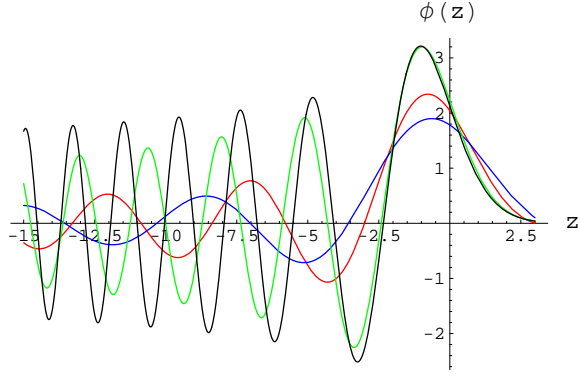}
\caption{\label{fign}The wave function $\phi(z)$ for $\beta=0$
(black line), $\beta=0.1$ (green line), $\beta=1$ (red line), and
$\beta=2$ (blue line). The eigenenergies are the minus of the zeros
of $\phi(z)$ and $\alpha=\hbar=1$. }
\end{center}
\end{figure}

In the presence of the minimal length, it is more appropriate to
study the problem in the momentum space. Because of the linear form
of the potential, the generalized Schr\"odinger equation can be cast
into a first-order differential equation in this space. Now we
define a new variable $\displaystyle z=x-\frac{E}{mg}$ and rewrite
the generalized Schr\"odinger equation (\ref{H0}) in the momentum
space as
\begin{eqnarray}
\tan^2\left(\sqrt{\beta}p\right)\phi(p)+2i\beta m^2g\hbar\phi'(p)=0,
\end{eqnarray}
where the prime denotes the derivative with respect to $p$. This
equation admits the following solution
\begin{eqnarray}
\phi(p)=\phi_0\exp\left[\frac{i\alpha}{\beta
\hbar}\left(\frac{\tan\left(\sqrt{\beta}p\right)}{\sqrt{\beta}}-p\right)\right].
\end{eqnarray}
where $\phi_0$ is a constant and $\alpha^{-1}=2m^2g$. To proceed
further and for the sake of simplicity, let us work in the units of
$\alpha=\hbar=1$. Now the energy eigenvalues are the minus of the
roots of the following wave function in the quasiposition space
representation
\begin{eqnarray}
\phi(z)=\phi_0\int_{-\frac{\pi}{2\sqrt{\beta}}}^{\frac{\pi}{2\sqrt{\beta}}}\exp\left[\frac{i}{\beta
}\left(\frac{\tan\left(\sqrt{\beta}p\right)}{\sqrt{\beta}}-p\right)\right]e^{ipz}\mathrm{d}p.
\end{eqnarray}
To evaluate the integral, it is more appropriate to use the new
variable $u=\tan\left(\sqrt{\beta}p\right)/\sqrt{\beta}$ and write
the wave function as
\begin{eqnarray}
\phi(z)=\phi_0\int_{-\infty}^{\infty}\frac{\left(1+i\sqrt{\beta}u\right)^{\frac{z-1/\beta}{\sqrt{\beta}}}}{\left(1+\beta
u^2\right)^{1+\frac{z-1/\beta}{2\sqrt{\beta}}}}e^{i\frac{u}{\beta}}\mathrm{d}u.
\end{eqnarray}
This integral can be evaluated numerically and the resulting
solutions for $\beta=\big\{0,\frac{1}{10},1,2\big\}$ are depicted in
Fig.~\ref{fign}. As the figure shows, the energy spectrum increases
in the presence of the minimal length i.e. $\beta\ne0$. The
calculated energy eigenvalues for the first ten states are also
shown in Table \ref{tab1} which confirms the positive shift in the
energy levels of the quantum bouncer in the GUP framework.

\begin{table}
\begin{center}
\begin{tabular}{ccc}\hline
$n$&\hspace{1cm}$\beta=0$  & \hspace{1cm}$\beta=1$ \\\hline
1& \hspace{1cm}2.33811     & \hspace{1cm}3.00000 \\
2& \hspace{1cm}4.08795     & \hspace{1cm}5.77871 \\
3& \hspace{1cm}5.52056     & \hspace{1cm}8.36567 \\
4& \hspace{1cm}6.78671     & \hspace{1cm}10.8569 \\
5& \hspace{1cm}7.94413     & \hspace{1cm}13.2878 \\
6& \hspace{1cm}9.02265     & \hspace{1cm}15.6765 \\
7& \hspace{1cm}10.0402     & \hspace{1cm}18.0333 \\
8& \hspace{1cm}11.0085     & \hspace{1cm}20.3649 \\
9& \hspace{1cm}11.9360     & \hspace{1cm}22.6761 \\
10& \hspace{1cm}12.8288     & \hspace{1cm}24.9702 \\\hline
\end{tabular}
\end{center}
\caption{\label{tab1}The first ten quantized energies of a bouncing
particle in GUP formalism for $\alpha=\hbar=1$.}
\end{table}

%**************************************************************************
To explain how the energy spectrum of the quantum bouncer increases
in the GUP scenario, let us study the dynamics of a particle in the
classical domain as
\begin{eqnarray}
\frac{dP}{dt}=\{P,H\},
\end{eqnarray}
where the Poisson bracket in classical mechanics corresponds to the
quantum mechanical commutator via
\begin{eqnarray}
\frac{1}{i\hbar}[\hat{A},\hat{B}]\Rightarrow\{A,B\}.
\end{eqnarray}
Now for the quantum bouncer we have
\begin{eqnarray}
\frac{dP}{dt}=mg\{P,X\}=-mg\left(1+\beta P^2\right)=-mG(P),
\end{eqnarray}
where $G(P)=g\left(1+\beta P^2\right)>g$ is the effective
gravitational acceleration. So an extra $P^2$ dependent
force $F'=-\beta mgP^2$ acts on the particle which increases its
energy in the presence of the minimal length.
%************************************************************************

It is worth to mention that the results of the Nesvizhevsky's
experiment have a quite large error bars. If we consider $\beta$ as
a universal constant, for example the value that Brau has found by
studying the hydrogen atom, we obtain \cite{r2,Brau}
\begin{eqnarray}
\beta<2\times 10^{-5}\,\,
\mathrm{fm}^2,\hspace{1cm}\mbox{and}\hspace{1cm}\Delta E_1\simeq
\Delta E_2 < 10^{-19} \,\,\mathrm{peV}.
\end{eqnarray}
These upper bounds tell us that the effects of the minimal
observable length are largely unobservable in the experiment, since
the maximal precision is $10^{-2} \,\,\mathrm{peV}$ \cite{293}.

In analogy with electrodynamics, the observation of spontaneous
decay of an excited state in Nesvizhevsky's  experiment could
demonstrates quantum behavior of the gravitational field as a
manifestation of the Planck-scale effect \cite{30}. Since the energy
spectrum of the quantum bouncer increases in the presence of the
minimal length, we expect that the rate of this decay changes as a
trace of quantum gravitational effects via existence of a minimal
length scale. The transition probability in the quadrupole
approximation and in the presence of GUP is \cite{pedram1,NP}
\begin{eqnarray}
\Gamma^{(\mbox{\footnotesize{GUP}})}_{k\longrightarrow
n}&=&\frac{512}{5}\frac{\left(\lambda^{(\mbox{\footnotesize{GUP}})}_{k}-
\lambda^{(\mbox{\footnotesize{GUP}})}_{n}\right)^5}{\left(\lambda_{k}-
\lambda_{n}\right)^{8}}\left(\frac{m}{M_{Pl}}\right)^{2}\frac{E^{5}_{0}c}{\gamma^{4}(\hbar
c)^{5}},
\end{eqnarray}
where $\gamma=\left(\alpha\hbar^2\right)^{-1/3}$, $E_{0}=mg/\gamma$,
$\lambda^{(\mbox{\footnotesize{GUP}})}_{n}=E_n/E_0$, and
$-\lambda_n$ are the zeros of the Airy function. Although this rate
is expected to be low, we hope to find the effects of the
generalized uncertainty principle on the transition rate of ultra
cold neutrons in the future more accurate experiments.

\subsection{Semiclassical approximation}
\begin{figure}
\begin{center}
\includegraphics[width=10cm]{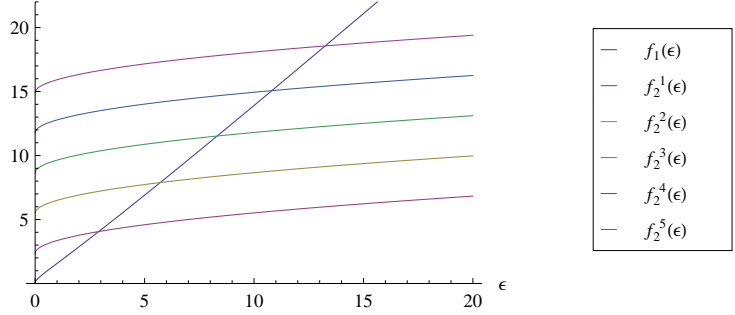}
\caption{\label{fig2}The schematic solutions of Eq.~(\ref{root}) for
$\beta=1$ and $\alpha=\hbar=1$. $f_1(\epsilon)=(\alpha+\beta
\epsilon)\arctan\sqrt{\beta \epsilon/\alpha}$ and
$f_2^n(\epsilon)=\sqrt{\alpha\beta
\epsilon}+\left(n-1/4\right)\pi\hbar\beta^{3/2}$.}
\end{center}
\end{figure}

We can also estimate the energy spectrum using the semiclassical
scheme. There are two options for writing the Hamiltonian of the
quantum bouncer $H=P^2/2m+mgX$ in the classical domain. The first
one is based on the KMM's proposal which results in
\begin{eqnarray}\label{sem1}
\frac{p^2}{2m}+mg(1+\beta p^2)x=E,
\end{eqnarray}
and the second one is based on Eqs.~(\ref{x0p01}) and (\ref{x0p02}),
i.e.,
\begin{eqnarray}\label{sem2}
\frac{\tan^2(\sqrt{\beta}p)}{2\beta m}+mgx=E.
\end{eqnarray}
Since both formulations are equivalent representations of the same
algebra, we expect that they result in the same energy spectrum. The
approximate energy eigenvalues can be obtained using Bohr-Sommerfeld
quantization rule
\begin{eqnarray}
\oint p\,\mathrm{d}x=
\left(n-\frac{1}{4}\right)h,\hspace{2cm}n=1,2,\ldots,
\end{eqnarray}
where the presence of the WKB corrected term $-1/4$ is due to the
fact that the wave function can penetrate into the right hand
classically forbidden region, but it is exactly zero for $x\leq0$.
Bohr-Sommerfeld formula for the first proposal (\ref{sem1}) results
in
\begin{eqnarray}
\oint
p\,\mathrm{d}x=2\int_0^{\epsilon}\sqrt{\frac{\epsilon-x}{\alpha+\beta
x}}\,\mathrm{d}x,
\end{eqnarray}
where $\epsilon=E^{(SC)}/mg$ is the rescaled
semiclassical energy. For the second representation (\ref{sem1}) we
have
\begin{eqnarray}
\oint
p\,\mathrm{d}x=\frac{2}{\sqrt{\beta}}\int_0^{\epsilon}\arctan\left(\sqrt{\frac{\beta}{\alpha}(\epsilon-x)}\right)\,\mathrm{d}x.
\end{eqnarray}
It is straightforward to check that both integrals are equivalent
and the energy eigenvalues are given by the roots of the following
equation
\begin{eqnarray}\label{root}
(\alpha+\beta \epsilon)\arctan\left(\sqrt{\frac{\beta
\epsilon}{\alpha}}\right)-\sqrt{\alpha\beta
\epsilon}-\left(n-\frac{1}{4}\right)\pi\hbar\beta^{3/2}=0.
\end{eqnarray}
To zeroth-order of the GUP parameter, this equation admits the
solution which is given by Eq.~(\ref{BS}), i.e.
$\epsilon^0_n=\left((9/4)\alpha\left[\pi
\hbar\left(n-1/4\right)\right]^2\right)^{1/3}$. Moreover, to the
first-order we have the following solution:
\begin{eqnarray}\label{root2}
\epsilon^1_n=\epsilon^0_n\left(1+\frac{2}{15}\frac{\beta}{\alpha}\epsilon^0_n\right).
\end{eqnarray}
Figure \ref{fig2} shows the schematic solutions of Eq.~(\ref{root})
for $\beta=1$ and $\alpha=1$. In Table \ref{tab2} we have reported
the exact and semiclassical energy spectrum of the quantum bouncer
in the GUP framework for $\beta=1$. As the results show, the
semiclassical solutions agree well with the exact ones up to a high
degree of accuracy of $\sim0.1\%-1\%$ even for the low energy
quantum states.

%*******************************************************************
\subsection{WKB approximation}
To check the validity of the Bohr-Sommerfeld quantization rule for
this modified quantum mechanics, let us write the first-order
generalized Schr\"odinger equation as
\begin{eqnarray}
-\frac{\hbar^2}{2m}\frac{\partial^2\psi(x)}{\partial x^2}+
\frac{\hbar^4}{3m}\frac{\partial^4\psi(x)}{\partial
x^4}+V(x)\psi(x)=E\,\psi(x),
\end{eqnarray}
and take
\begin{eqnarray}
\psi(x)=e^{i\Phi(x)},
\end{eqnarray}
where $\Phi(x)$ can be expanded as a power series in $\hbar$ in the
semiclassical approximation, i.e.,
\begin{eqnarray}
\Phi(x)=\frac{1}{\hbar}\sum_{n=0}^{\infty}\hbar^n\Phi_n(x).
\end{eqnarray}
So we have
\begin{eqnarray}
\frac{\partial^2\psi(x)}{\partial
x^2}&=&-\left(\Phi'^2-i\Phi''\right)\psi(x),\\
\frac{\partial^4\psi(x)}{\partial x^4}&=&\left(\Phi'^4
-6i\Phi'^2\Phi''-3\Phi''^2-4\Phi'''\Phi' +i \Phi''''\right)\psi(x),
\end{eqnarray}
where $\Phi'$ indicates the derivative of $\Phi$ with respect to
$x$. To zeroth-order ($\Phi(x)\simeq\Phi_0(x)/\hbar$) and for
$\hbar\rightarrow0$ we obtain
\begin{eqnarray}
\Phi_0'^2+\frac{2}{3}\beta\Phi_0'^4=2m\left(E-V(x)\right).
\end{eqnarray}
Now the comparison with Eq.~(\ref{eqP}) shows $\Phi_0'=p$ and
consequently
\begin{eqnarray}
\psi(x)\simeq \exp\left[{\frac{i}{\hbar}\int p\,\mathrm{d}x}\right],
\end{eqnarray}
which is the usual zeroth-order WKB wave function obeying the
Bohr-Sommerfeld quantization rule. The generalization of this result
to higher orders perturbed Hamiltonians and the nonperturbative
Hamiltonian (\ref{H0}) is also straightforward. Indeed, the
agreement between the exact and semiclassical results is the
manifestation of the validity of the Bohr-Sommerfeld quantization
rule in this modified quantum mechanics.

\subsection{Perturbative study}
Now, following Ref.~\cite{Brau}, let us study this problem
perturbatively. To first-order of the GUP parameter the Hamiltonian
can be written as
\begin{eqnarray}\label{eqP}
H=\frac{p^2}{2m}+\beta \frac{p^4}{3m}+V(x)=H_0+\beta \frac{p^4}{3m}.
\end{eqnarray}
Using the first order time-independent perturbation theory, the
energy spectrum is given by
\begin{eqnarray}
E_n^1\simeq E_n^0+\Delta E_n,
\end{eqnarray}
where to the first order we have
\begin{eqnarray}
\Delta E_n&=&\frac{\beta}{3m}\langle p^4\rangle=\frac{4}{3}\beta m\langle (H_0-V)^2\rangle,\nonumber\\
&=&\frac{4}{3}\beta m\Big[(E_n^0)^2-2E_n^0\langle
V(x)\rangle+\langle V(x)^2\rangle\Big].\label{per}
\end{eqnarray}
Since $V(x)\sim x$ is a power-law potential, the virial theorem
gives
\begin{eqnarray}
\langle V(x)\rangle=\frac{2}{3}E_n^0.
\end{eqnarray}
So using $\langle
V(x)^2\rangle=\frac{8}{15}(E_n^0)^2$ \cite{Brau}, the relation (\ref{per}) reduces to
\begin{eqnarray}
\Delta E_n=\frac{4}{15}\beta (E_n^0)^2.
\end{eqnarray}
If we define ${\cal E}_n= E_n/mg$ we obtain
\begin{eqnarray}
{\cal E}_n^1\simeq {\cal E}_n^0
\left(1+\frac{2}{15}\frac{\beta}{\alpha}{\cal E}_n^0\right),
\end{eqnarray}
which has the form of the semiclassical solution (\ref{root2}) and
${\cal E}_n^0$ is given by \cite{Brau}
\begin{eqnarray}
{\cal E}_n^0 =\alpha^{1/3}\lambda_n,
\end{eqnarray}
As stated before, ${\cal E}_n^0$ are nearly equal to $\epsilon^0_n$
even for the lowest quantum state. Therefore, our results agree with
those of Ref.~\cite{Brau} up to the first order of the GUP
parameter.

\begin{table}
\begin{center}
\begin{tabular}{cccc}\hline
$n$&\hspace{1cm}$\epsilon_n^{\mathrm{exact}}$  &
\hspace{1cm}$\epsilon_n^{\mathrm{SC}}$& \hspace{1cm}$|\frac{\Delta
\epsilon_n}{\epsilon_n}|$
\\\hline
1&  \hspace{1cm}3.00000     & \hspace{1cm}2.90366 & \hspace{1cm} 0.032    \\
2&  \hspace{1cm}5.77871     & \hspace{1cm}5.71575 & \hspace{1cm} 0.011    \\
3&  \hspace{1cm}8.36567     & \hspace{1cm}8.31537 & \hspace{1cm} 0.006    \\
4&  \hspace{1cm}10.8569     & \hspace{1cm}10.8137 & \hspace{1cm} 0.004    \\
5&  \hspace{1cm}13.2878     & \hspace{1cm}13.2495 & \hspace{1cm} 0.003    \\
6&  \hspace{1cm}15.6765     & \hspace{1cm}15.6416 & \hspace{1cm} 0.002    \\
7&  \hspace{1cm}18.0333     & \hspace{1cm}18.0010 & \hspace{1cm} 0.002    \\
8&  \hspace{1cm}20.3649     & \hspace{1cm}20.3348 & \hspace{1cm} 0.001    \\
9&  \hspace{1cm}22.6761     & \hspace{1cm}22.6477 & \hspace{1cm} 0.001    \\
10& \hspace{1cm}24.9702     & \hspace{1cm}24.9433 & \hspace{1cm} 0.001    \\\hline
\end{tabular}
\end{center}
\caption{\label{tab2}The exact and semiclassical energy levels of
the quantum bouncer for $\beta=1$ and $\alpha=\hbar=1$.}
\end{table}

\section{Conclusions}
In this paper, we have investigated the effects of the generalized
uncertainty principle on the energy spectrum of a quantum bouncer.
Using two exact and equivalent nonperturbative representations, we
found the generalized Schr\"odinger equation in the momentum space
and obtained the corresponding energy eigenvalues and the
eigenfunctions. The second representation, in contrary to the KMM
representation, was formally self-adjoint and preserved the ordinary
nature of the position operator. Then, using the proper
semiclassical approximation, we found the almost accurate energy
spectrum even for the low lying eigenstates and similar to the
perturbative representations we observed a positive shift in the
energy levels. Also, we explicitly found the first order quantum
mechanical energy spectrum and showed the validity of the
Bohr-Sommerfeld quantization rule in this modified quantum
mechanics.


\begin{thebibliography}{99}
\bibitem{12}         D. Amati, M. Ciafaloni, and G. Veneziano, Phys. Lett. B \textbf{216} (1989) 41.
\bibitem{13}         M. Maggiore, Phys. Lett. B \textbf{304} (1993) 65.
\bibitem{14}         M. Maggiore, Phys. Rev. D \textbf{49} (1994) 5182.
\bibitem{15}         M. Maggiore, Phys. Lett. B \textbf{319} (1993) 83.
\bibitem{16}         L. J. Garay, Int. J. Mod. Phys. A \textbf{10} (1995) 145.
\bibitem{r2}         F. Brau, J. Phys. A \textbf{32} (1999) 7691.
\bibitem{17}         F. Scardigli, Phys. Lett. B \textbf{452} (1999) 39.
\bibitem{18}         S. Hossenfelder et al., Phys. Lett. B \textbf{575} (2003) 85.
\bibitem{19}         C. Bambi and F. R. Urban, Class. Quant. Grav. \textbf{25} (2008) 095006.
\bibitem{101}        K. Nozari and B. Fazlpour, Gen. Relativ. Grav. \textbf{38} (2006) 1661.
\bibitem{102}        B. Vakili, Phys. Rev. D \textbf{77} (2008) 044023.
\bibitem{103}        M. V. Battisti and G. Montani, Phys. Rev. D \textbf{77}  (2008) 023518.
\bibitem{104}        B. Vakili and H. R. Sepangi, Phys. Lett. B \textbf{651} (2007) 79.
\bibitem{105}        M. V. Battisti and G. Montani, Phys. Lett. B \textbf{656} (2007) 96.
\bibitem{106}        K. Nozari and T. Azizi, Gen. Relativ. Gravit. \textbf{38} (2006) 735.
\bibitem{7}          K. Nozari and P. Pedram, Europhys. Lett. {\bf 92} (2010) 50013, arXiv:1011.5673.
\bibitem{Nouicer}    K. Nouicer, J. Phys. A \textbf{39} (2006) 5125.
\bibitem{21}         J. Magueijo and L. Smolin, Phys. Rev. Lett. \textbf{88} (2002) 190403.
\bibitem{22}         J. Magueijo and L. Smolin, Phys. Rev. D \textbf{71} (2005) 026010.
\bibitem{23}         G. Amelino-Camelia, Int. J. Mod. Phys. D \textbf{11} (2002) 35.
\bibitem{24}         G. Amelino-Camelia, Phys. Lett. B \textbf{510} (2001) 255.
\bibitem{3}          J. L. Cortes and J. Gamboa, Phys. Rev. D \textbf{71} (2005) 065015.
\bibitem{main}       A. F. Ali, S. Das, and E. C. Vagenas, Phys. Lett. B \textbf{678} (2009) 497.
\bibitem{main2}      S.~Das, E.C.~Vagenas, and A.F.~Ali, Phys.~Lett.~B \textbf{690} (2010) 407.
\bibitem{pedram}     P. Pedram, Euro. Phys. Lett. \textbf{89} (2010) 50008, arXiv:1003.2769.
\bibitem{pedram3}    P.~Pedram, Phys.~Lett.~B \textbf{702} (2011) 295.
\bibitem{pedram1}    P.~Pedram, K.~Nozari, and S.H.~Taheri, JHEP \textbf{03} (2011) 093, arXiv:1103.1015.
\bibitem{Brau}       F. Brau and F. Buisseret, Phys. Rev. D \textbf{74} (2006) 036002, arXiv:hep-th/0605183.
\bibitem{NP}         K.~Nozari and P.~Pedram, Europhys.~Lett.~\textbf{92} (2010) 50013, arXiv:1011.5673.
\bibitem{Kempf}      A.~Kempf, G.~Mangano, and R.~B.~Mann, Phys.~Rev.~D \textbf{52} (1995) 1108, arXiv:hep-th/9412167.
\bibitem{main1}      S.~Das and E.C.~Vagenas, Phys.~Rev.~Lett.~\textbf{101} (2008) 221301, arXiv:0810.5333.
\bibitem{pedram2}    P. Pedram, Int. J. Mod. Phys. D \textbf{19} (2010) 2003, arXiv:1103.3805.
\bibitem{291}        V.V.~Nesvizhevsky et al., Nature \textbf{415} (2002) 297.
\bibitem{292}        V.V.~Nesvizhevsky et al., Phys.~Rev.~D \textbf{67} (2003) 102002.
\bibitem{293}        V.V.~Nesvizhevsky et al., Eur.~Phys.~J.~C \textbf{40} (2005) 479.
\bibitem{30}         G.~Pignol, K.V.~Ptotasov, and V.V.~Nesvizhevsky, Class.~Quant.~Grav.~\textbf{24} (2007) 2439.
\end{thebibliography}
\end{document}